\documentclass[aps,prl,twocolumn,groupedaddress,showpacs]{revtex4}
\usepackage{graphicx}
\usepackage{epstopdf}

\newcommand{\delete}[1]{}

\newcommand{\be}{\begin{equation}}
\newcommand{\ee}{\end{equation}}

\newcommand{\eff}{\text{eff}}

\def\beq{\begin{equation}}
\def\eeq{\end{equation}}
\def\bea{\begin{eqnarray}}
\def\eea{\end{eqnarray}}
\def\ba{\begin{array}}
\def\ea{\end{array}}

\begin{document}

\preprint{Stanford-ITP-03-17}

\title{Testing Atom and Neutron Neutrality with Atom Interferometry}


\author{Asimina Arvanitaki$^{1,2}$}
\email[]{aarvan@stanford.edu}
\author{Savas Dimopoulos$^1$}
\email[]{savas@stanford.edu}
\author{Andrew A. Geraci$^{1}$}
\email[]{aageraci@nist.gov}
\author{Jason Hogan$^1$}
\email[]{hogan@stanford.edu}
\author{Mark Kasevich$^1$}
\email[]{kasevich@stanford.edu}
\affiliation{$^1$Department of Physics, Stanford University, Stanford, CA
94305} \affiliation{$^2$Theory Group, Stanford Linear Accelerator Center,
Menlo Park, CA 94025}


\date{\today}

\begin{abstract}
We propose an atom-interferometry experiment based on the scalar Aharonov-Bohm effect which detects an atom charge at the $10^{-28}e$ level, and improves the current laboratory limits by 8~orders of magnitude. This setup independently probes neutron charges down to $10^{-28}e$, 7 orders of magnitude below current bounds.

\end{abstract}

\maketitle


{\it{Introduction}.} Charge quantization and atom neutrality in the
Standard Model (SM) are mysteries which are automatically solved when the theory is embedded in a Grand Unified group. Even then, Witten has shown \cite{Witten:1979ey} that in the presence of CP non-conservation magnetic monopoles acquire an electric charge that is proportional to the amount of CP violation, a non-quantized quantity. This suggests that we have to rethink our notion of atom neutrality even in the presence of a unifying group.

The first experiments to test charge cancelation between the
constituents of the atom came at the turn of the twentieth century \cite{Unni:2004}. These experiments placed a bound on $\frac{e+p}{e}$ of $10^{-21}$, a value that is only an order of magnitude
larger than the bound set by recent experiments \cite{Unni:2004, Eidelman:2004wy}. Experiments to detect individual neutron charges independently required different technology, took longer to develop, and have eventually reached a sensitivity of $10^{-21}e$, similar to that of atom neutrality experiments \cite{Unni:2004, Eidelman:2004wy}. Over eighty years after the first precision experiment on atom neutrality
was performed, atom interferometry pushes the precision frontier, and
provides a new tool for testing fundamental physics by measuring effects on
individual atoms \cite{berman}. An experiment to test the equivalence
principle and modifications of gravity is already under construction
\cite{Dimopoulos:2006nk}. In this Letter, we propose a modification of that
experiment based on the scalar Aharonov-Bohm effect
\cite{Aharonov:1959fk} that can detect atom and, independently, neutron charges down to $10^{-28}e$. 

Because of the topological nature of the Aharonov-Bohm effect, the atoms
are under the influence of pure gauge electromagnetic potentials; all electric and magnetic fields are zero. As a result, there are ideally no forces acting on the atoms, and systematics from the finite atom polarizability are avoided. If the atom carries a small charge $\epsilon e$, its wave-function will acquire a phase $\frac{\epsilon e}{\hbar} V t$, where $e$ is the electron
charge, $t$ is the time spent in the region of electric potential $V$, and
$\epsilon$ is the ratio of atom charge compared to the electron charge.

We begin with the experimental setup, and analyze possible systematics as
well as the ultimate sensitivity of the experiment with different upgrades.
We end with a discussion on the theoretical motivation behind
infinitesimally charged atoms.

\begin{figure}[!t]
\begin{center}
\includegraphics[width=2.0in]{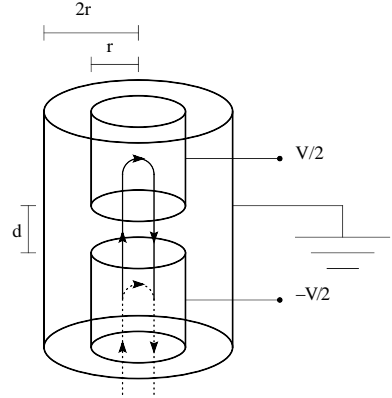}
\caption{The experimental set-up including
particle trajectories. The upper and lower electrode tubes are operated at
potentials $V/2$ and $-V/2$ respectively, as described in the text.}
\label{Fig:Experiment}
\end{center}
\end{figure}
\begin{figure}[!t]
\begin{center}
\includegraphics[width=3.0in]{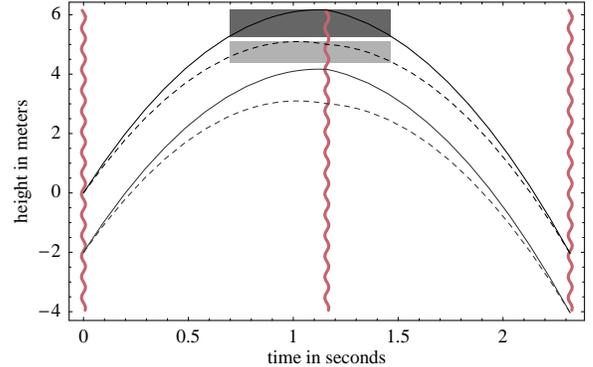}
\caption{The atom wave-function trajectories as a function of time. The
areas in dark and light gray indicate when the fast and slow atom wavefunctions are traveling in a region of voltage V/2 and -V/2, respectively. Below are shown the trajectories for a second atom cloud launched at the same time, in order to control systematics from laser phase noise.} \label{Fig:Trajectory}
\end{center}
\end{figure}

{\it{Experimental Setup}.} The proposed apparatus is based on a 10 m interferometer designed
to test the equivalence principle \cite{Dimopoulos:2006nk}. Evaporatively
cooled $^{87}$Rb atoms are launched vertically with an initial velocity
$v_L\sim 10~ \frac{\text{m}}{\text{s}}$. A series of laser pulses ($\frac{\pi}{2}-\pi-\frac{\pi}{2}$ sequence) acts as beamsplitters and mirrors for the atoms, splitting the atom
wave-function into a superposition of space-time trajectories with momentum difference $\hbar k_{\eff}$, and then recombining them in order to interfere. 
This momentum difference also sets the maximum
spatial separation of the wave-packets at the maxima of their trajectories.  Within the next few years, the application of Large Momentum Transfer (LMT) beamsplitters \cite{Denschlag:2002, Mcguirck:2000, Weitz:1994} in atom interferometry will likely be realized. With LMT, a velocity splitting of
$\frac{\hbar k_{\eff}}{m_{\text{\tiny{atom}}}}\sim 1~\frac{\text{m}}{\text{s}}$ may be possible and would result in a
separation of $1.07$~m between the fast and the slow wave-function component at the highest points
of their trajectories.

Taking advantage of this large spatial separation, we can introduce regions
of potential $V/2$ and $-V/2$ in the trajectories of the fast and slow
components respectively, as illustrated in Fig. \ref{Fig:Experiment}. Both
the fast and slow trajectories extend vertically along the axis of the
lower cylindrical electrode of radius $r$. Only the fast-component
trajectory extends upwards into a second cylindrical
electrode of radius $r$.  An axial gap $d$ separates the two cylindrical
electrodes, and the trajectory of the slow-component does not extend into
the gap region. Assuming $d\sim$ cm, strong electric fields of order $V/d\sim 10^7\frac{\text{V}}{\text{m}}$ are present in the gap. To avoid these, the voltage
is applied only when the atoms are well inside the electrode tubes. 

{\it{Sensitivity.}} If the atom carries a small charge $\epsilon e$, the
phase difference between the fast and slow component of the atom wave-function is:
\be \epsilon e \int \frac{V} {\hbar} dt   \label{phase}. \ee For $V=10^5$
Volts and an interaction time of 0.7 s, the phase shift becomes $\sim
10^{20} \epsilon$. With $10^6$ atoms, assuming shot-noise limited phase sensitivity of
$10^{-3}$ rad per trial and integrating over $10^6$ trials, the experiment
can probe phase shifts of $10^{-6}$ rad and measure atom charges down to
$\epsilon \sim 10^{-26}$. These bounds are usually expressed in terms of the average charge per nucleon, $\eta \equiv \frac{\epsilon}{A}$, where $A$ is the total number of nucleons in the atom.
In this language, the experimental reach is $\eta \sim 10^{-28}$. The current laboratory limit is
$\eta = 10^{-22}$ \cite{Eidelman:2004wy}.

Future prospects for atom interferometry involve increasing the number of atoms per trial to $10^7-10^8$ and/or using entangled states of atoms. In experiments with entangled atoms, the atom phases add coherently in each shot and the sensitivity becomes Heisenberg-limited \cite{heisenberg}. These prospects combined could allow an improvement of at least two orders of magnitude, bringing the
experimental reach down to $\eta \sim 10^{-30}$.  

A measurement of $\epsilon$ determines a linear combination of the proton, electron and neutron charges. An independent bound on the charge of the neutron can be placed by performing a differential measurement between $^{87}$Rb and $^{85}$Rb atoms in the same atom cloud. In this case, the ultimate sensitivity is $10^{-28} e$, an approximate 7 orders of magnitude improvement on current bounds. This experiment, combined with the measurement of the individual Rb atom charges, will give independent measurements of the neutron charge and the sum of the proton and electron charges. Measuring the charges of different atoms will improve the neutron charge measurement, but is always sensitive to the sum of the proton and electron charges.

To obtain the desired experimental sensitivity, other stochastic sources of
interferometer phase noise must be kept below the fundamental atom noise limit.
Examples of such sources include laser phase noise and the fluctuating
initial positions $z_i$ of the atomic clouds, which couple to gravity
gradients $T_{zz}$ and lead to a phase shift $-k_{\rm{eff}} T^2 T_{zz} z_i
\approx 7 \frac{\text{rad}}{\text{mm}}$, where $T=1.16$ s is the interrogation time of the experiment, the time interval between the laser pulses in the $\frac{\pi}{2}-\pi-\frac{\pi}{2}$ sequence.  To suppress laser phase noise, we consider operating a second interferometer in the same tube, vertically separated from the primary interferometer by approximately 2 meters to avoid the high voltage
electrodes, and subject to the same laser pulses. We then compare the
differential phase shift between the two interferometers as a function of the tube
voltage in the primary interferometer;  laser phase noise is the same for both interferometers and cancels as common mode. For the neutron charge measurement, the differential measurement is between the two collocated Rb isotopes and there is no need for the additional interferometer. Assuming the local gravitational gradient can be reduced to 10$\%$ by an engineered local mass distribution,
the initial position fluctuations between the interferometers must be
controlled at the $1~ \mu$m level to allow shot-noise-limited sensitivity,
and at the 10 nm level for a Heisenberg-limited interferometer.  Also,
variations in the initial launch velocity can contribute to the noise
through a gravity gradient phase shift $-k_{\rm{eff}} T^3 T_{zz} v_L \approx
10^4$ rad. The launch velocity $v_L$ must therefore be maintained
consistent at the level of 1 $\frac{\mu \text{m}}{\text{s}}$ for shot-noise
limited sensitivity and 10 $\frac{\text{nm}}{\text{s}}$ to reach the Heisenberg-limit. 

{\it{Systematics}.}  
The potential is a control parameter that distinguishes the
scalar Aharonov-Bohm effect from other systematic phase shifts.
Assuming stochastic noise sources can be controlled at the shot-noise
level, only systematic phase shifts that depend on the voltage can potentially limit sensitivity. Such an effect comes from the electric field gradient near the openings of the tube electrodes, which induces a dipole force on the atoms. To reduce this effect below detection, the voltage is turned on when the atoms are at least 10 radii inside the tube electrodes. Since the atoms spend most of their time at the highest point of their trajectories, this procedure does not significantly affect the experimental sensitivity. 

Turning the potential on and off involves transient currents which result in transient magnetic fields.
We consider a linear voltage ramp to $V_0=100$ kV
in a time $\tau=0.1$ s. For a 1-meter long electrode of radius $r=1$ cm surrounded by the
coaxial grounded tube of radius $2r$, the transient current flowing during
the charging process is approximately $8 \times 10^{-5}$ A. Any asymmetry
in the transient currents creates a magnetic field inside the tube.
Assuming complete asymmetry the magnitude of the field along the axis of the tube is at most
$  B_{tr}\sim \frac{\mu_0}{2 r}I_{tr} \sim 50 ~\mu \text{G}. $
Since the atoms are in the $m_F=0$ state along the z axis, they interact to second order with magnetic fields along z:
$ -\frac{1}{2} \alpha_m \vec{B}^2 = -\frac{1}{2} \alpha_m (\vec{B}_{tr}+\vec{B}_0)^2, $
where $\vec{B}_0=(1~\text{mG})~ \hat{z}$ is a constant bias magnetic field used to define the axis of quantization for the atoms. 
The transient field produces an additional acceleration along the direction of motion: \be a=
\alpha_m \frac{\partial_z (2 \vec B_{\rm{tr}}\cdot  \vec
B_0+\vec {B^2}_{\rm{tr}} )}{m}\sim 10^{-11}
\rm{\frac{\text{m}}{\text{s}^2}}, \ee assuming significant variation of the transient
field over 1 cm and  $\alpha_m = 2 \pi \times 575\frac{\text{Hz}}{\text{G}^2}$ for $^{87}$Rb.

One of the advantages of the aforementioned voltage-symmetric ($V/2$ and $-V/2$) design is that, as long as both the fast and slow atoms experience similar transient field gradients,
the induced magnetic phase shift will partially cancel. To make a
quantitative estimate, we separately consider the following set of magnetic
fields and their gradients: the background bias field $B_0$, the additional
transient magnetic fields present in the upper $(V/2)$ and lower $(-V/2)$
electrodes while the voltage is being ramped up ($B_{t1}$ and $B_{t2}$,
respectively), and the additional transient magnetic field present in the
upper and lower electrodes while the voltage is being brought back to
ground ($B_{t3}$ and $B_{t4}$, respectively).

To properly estimate the transient field effects we conduct the full phase
calculation, as \cite{Bongs}: \be \Delta \phi_{\rm{total}} =
\Delta \phi_{\rm{propagation}} + \Delta \phi_{\rm{laser}} + \Delta
\phi_{\rm{separation}}. \ee  The primary voltage-dependent contributions to
the phase shift are listed in Table \ref{magterms}.  
Terms of order $10^{-4}$ rad may persist even with perfect upper-lower
symmetry and equal ramp times. These terms may be present even when using two collocated Rb isotopes, since their respective $\alpha_m$ values are significantly different.

\begin{table}[!t]
\begin{center}
  \begin{tabular}{@{}|c|c|c|@{}}
  \hline \hline
  Aharonov-Bohm Signal & phase shift(rad) & Scaling \\
  \hline
  $\epsilon e V (t_{\rm{off}}-t_{\rm{on}}) / \hbar$ & $10^{20} \epsilon$
  & $V$\\
  \hline \hline
    Systematic & phase shift(rad) & Scaling \\
    \hline
    Magnetic (symmetric) &  &  \\
    \hline
    $-\frac{1}{3 \hbar} g t_{\rm{off}} v_L \alpha_m \tau_2^3 (\frac{\partial{B_{t4}}}{\partial{z}})^2$ &  $-1 \times 10^{-3}$ & $V^2\tau_2$ \\
    $\frac{1}{3 \hbar} g t_{\rm{off}} v_L \alpha_m \tau_2^3 (\frac{\partial{B_{t3}}}{\partial{z}})^2$ & $1 \times 10^{-3}$ & $V^2\tau_2$ \\
    $\frac{1}{6 \hbar} g^2 t_{\rm{off}}^2 \alpha_m \tau_2^3 (\frac{\partial{B_{t4}}}{\partial{z}})^2$ & $7 \times 10^{-4}$ & $V^2\tau_2$ \\
    $-\frac{1}{6 \hbar} g^2 t_{\rm{off}}^2 \alpha_m \tau_2^3 (\frac{\partial{B_{t3}}}{\partial{z}})^2$ & $-7 \times10^{-4}$ &
    $V^2\tau_2$
    \\
    $-\frac{1}{3 \hbar} g t_{\rm{on}} v_L \alpha_m \tau_1^3 (\frac{\partial{B_{t2}}}{\partial{z}})^2$ &  $-5 \times 10^{-4}$ & $V^2\tau_1$   \\
    $\frac{1}{3 \hbar} g t_{\rm{on}} v_L \alpha_m \tau_1^3 (\frac{\partial{B_{t1}}}{\partial{z}})^2$ & $5 \times
    10^{-4}$ & $V^2\tau_1$
    \\

    \hline
    Magnetic (non-symmetric) & & \\
    \hline
    $-\frac{1}{3 m} g k_{\rm{eff}} (t_{\rm{off}}-t_{\rm{on}}) \alpha_m \tau_1^3 (\frac{\partial{B_{t1}}}{\partial{z}})^2$ &  $-5 \times 10^{-5}$ & $V^2\tau_1$ \\
    $-\frac{1}{m} B_0 k_{\rm{eff}} \alpha_m \tau_1^2 \frac{\partial{B_{t1}}}{\partial{z}}$ &  $-9 \times 10^{-5}$ & $V B_0 \tau_1$ \\
    \hline
    Electric polarizability &  $10^{-14}$  & $V^2$ \\
    \hline \hline
  \end{tabular}
\caption{\label{magterms} Estimated voltage-dependent signal and systematic
phase shifts. V is the voltage applied on the tubes. $t_{\rm{on}}$ and
$t_{\rm{off}}$ are the times when the voltage is turned on and off,
respectively. The magnetic systematics are divided into terms that vanish
for perfect upper-lower symmetry ($\partial_z B_{t1}=\partial_z B_{t2}$, $\partial_z B_{t3}=\partial_z B_{t4}$) and
equal ramping times ($\tau_{1}=\tau_{2}$) and terms that do not.}
\end{center}
\end{table}

Both the symmetric and non-symmetric terms are suppressed by
reducing the ramp times $\tau_i$, and improving the symmetry of the
electrode geometry so that the transient fields $B_{ti}$ and their
gradients are reduced.  The transient magnetic fields vanish for
current flow uniform on the tube surface and parallel to the tube axis during the charging process. In our estimate appearing in
Table \ref{magterms} we have maximally exaggerated the asymmetry to
demonstrate the most conservative case. 
We estimate that it is possible to reduce asymmetries in the current
flow as well as the geometry of the configuration to the level of a few
percent. In any case, even though the systematic effects of transient
fields may be significant, we can remove them from the analysis using their scaling with experimental control parameters. Most of these systematic contributions are proportional to $\vec B_{\rm{tr}}^2$ and scale with
$V^2$ and ${\tau}$, which is different from the Aharonov-Bohm effect scaling. The only systematic effect which scales linearly with $V$ also depends on $\vec B_0$ and can therefore be removed from the analysis as well. Finally, an inner layer of magnetic field shielding can be used to further mitigate these effects.

There is another potential source of voltage dependent systematics: the walls of the high-voltage tube electrode experience a small deformation that depends on voltage due to electrostatic pressure. This deformation affects the diffraction of the laser from the walls of the tube and creates a small spurious voltage dependent phase shift. This systematic can be pushed below the Heisenberg statistics sensitivity level, by reducing the laser beam waist below 5 mm for an electrode tube of 1 cm radius.

An earlier proposal  to test matter neutrality \cite{Champenois} already employs the scalar Aharonov-Bohm effect. The apparatus is based on a Mach-Zehnder atom interferometer and the design sensitivity is $\eta \sim \frac{10^{-21}}{\sqrt{\text{Hz}}}$, compared to $\eta \sim \frac{10^{-27}}{\sqrt{\text{Hz}}}$ for the current proposal.

{\it{Theoretical motivation}.} This experiment's thirty decimal reach pushes precision measurements in a new regime and may probe effects of Planck scale physics. But is there a framework that predicts tiny atomic charges, while maintaining the success of  gauge coupling unification and approximate charge quantization? Recent developments suggest that topological shifts of the SM electric charges, similar to the Witten effect \cite{Witten:1979ey}, occur when ordinary particles carry quantum hair under massive higher spin fields\cite{Dvali}. For example, if the electron carries a magnetic charge under these fields and they have a CP violating mixing to EM, the electron charge is shifted by an amount proportional to its magnetic charge and the CP violation. If the higher spin fields are in the bulk of an extra dimensional model while the SM particles are localized on the brane boundary, these fields' couplings to the SM get volume diluted and the resulting charge shifts are naturally small.

Violations of charge quantization are also expected when the photon has a small mass, $m_{\gamma}$.   If the photon is massive, SM charges no longer need to satisfy gauge anomaly cancellation relations. The charge shift should be proportional to $\frac{m_{\gamma}}{M}$, where $M$ is the cutoff for the massive photon theory. If the scale $M$ is identified with the weak scale, then photon masses down to $m_\gamma = 10^{-19}~\text{eV}$ can be probed, 4 orders of magnitude below present limits \cite{Adelberger:2003qx}.

In a more conventional framework, the charges of ordinary particles can be
shifted by a small family-universal amount with all gauge anomalies cancelled (in the presence of right-handed neutrinos), if these shifts are proportional to the particles' Baryon-Lepton (B-L) number: all baryon EM charges are shifted by $\epsilon$ and all lepton EM charges by $-\epsilon$.
As a  result, the proton and the electron still have equal and opposite
charges, while the neutron and the neutrino acquire small opposite charges. This is dynamically realized when B-L is an
extra SM U(1) gauge symmetry  that is broken at high energies by a
Higgs $\Phi$ that has a small hypercharge $\epsilon$. The EM photon acquires a small component of the B-L gauge boson, and the particle EM charges
are shifted as described above by an amount proportional to $\epsilon$. 

In these models gauge coupling unification is accommodated in a deconstructed group, such as $SU(5)\times SU(3)\times SU(2)\times U(1)_{Y'} \times U(1)_{B-L}$ \cite{Weiner:2001pv, Csaki:2001qm}. In this case, $SU(5)\times SU(3)\times SU(2)\times U(1)_{Y'} $ is broken diagonally to the SM gauge group. If the $SU(5)$ gauge coupling is much weaker than  those of the $SU(3)\times SU(2)\times U(1)_{Y'}$ group, the Grand Unified values of the SM gauge couplings is determined by the $SU(5)$ gauge coupling and unification is preserved. $\Phi$ has charge $\epsilon$ under $U(1)_{Y'}$ and avoids the forbidding constraints from charge quantization in a non-abelian group. More minimal deconstructed groups, such as $SU(5)\times U(1)_{Y'}\times U(1)_{B-L}$, are also possible.

A higher dimensional realization of this scenario puts $SU(5)\times U(1)_{Y'}\times U(1)_{B-L}$ in the bulk. When $U(1)_{Y'}$ is localized in the extra dimension \cite{Batell:2005wa} close to the SM brane, its coupling will appear large compared to that of the $SU(5)$, that remains flat and gets diluted by volume effects. In addition, if $\Phi$ lies away from the localized $U(1)_{Y'}$  gauge symmetry, its Y' charge will be exponentially suppressed, dynamically explaining the smallness of $\epsilon$.

The above examples are under study \cite{atomcharges}, and show that atom neutrality should not be taken for granted. Charge quantization in the SM model can be violated in a way that preserves gauge coupling unification. Detection of an infinitesimal atom charge would be a powerful indication of new physics far above the TeV scale.



{\it{Acknowledgements}} We would like to thank G. Dvali for valuable discussions. A.A. would also like to thank T. Gherghetta, P. W. Graham, J. March-Russell and S. Rajendran for useful discussions.

\end{document}